# Approximate Black Holes for Numerical Relativity

Maurice H.P.M. van Putten *

CRSR & Cornell Theory Center

Cornell University

Ithaca, NY 14853-6801 †

(July 30, 1996)

## Abstract

Spherically symmetric solutions in Brans-Dicke theory of relativity with zero coupling constant, $\omega = 0$, are derived in the Schwarzschild line-element. The solutions are obtained from a cubic transition equation with one small parameter. The exterior space-time of one family of solutions is arbitrarily close to the exterior Schwarzschild space-time. These nontopological solitons have some similarity with soliton stars, and are proposed as candidates for *approximate black holes* for the use in numerical relativity, in particular for treatment of horizon boundary conditions.

---

*Electronic address: `vanputte@spacenet.tn.cornell.edu`

†Present address: Department of Mathematics, MIT, Cambridge, MA 02139.

# I. INTRODUCTION

Numerical relativity is considered the prime vehicle for the prediction of gravitational wave forms in the late stages of coalescence in binary systems of black holes and neutron stars. This effort is complementary to recent analytical approaches for the prediction of waveforms in the early stages of coalescence [2,1]. Prediction of waveforms is an important factor in the detection of gravitational radiation by LIGO and VIRGO [6,5]. The greatest challenges in numerical relativity involve the simulations of multiple orbits in the presence of black holes or horizon formation. One of the open problems is proper numerical treatment of horizon boundary conditions.

In this *Letter*, we explore the possibility of using *approximate black holes* (ABHs) for the purpose of numerical relativity by introducing a small parameter, $\epsilon$, in a modified setting of the physics of the problem. The ABHs to be considered are nontopological solitons in that they are produced by a real scalar field in an asymptotically flat space-time, such that the exterior space-time of ABHs approximates the exterior space-time of black holes (BHs) as $\epsilon \to 0$. As a first step in this direction, the existence of such solitons shall be demonstrated in the Schwarzschild line-element, leaving the general case as an open problem. The fact that the scalar field is real, and that the stress-energy tensor in Brans-Dicke does *not* stem from an additive contribution to the Lagrangian [1], however, makes these objects distinct from existing soliton stars.

Concentrating on the problem of black hole coalescence, and thereby ingoring the presence of matter, we are at liberty to consider Brans-Dicke theory with $\omega = 0$ as an approximation of general relativity. As will be clarified further in what follows, the limit of small $\epsilon$, instead of large $\omega$, will serve to approximate general relativity in a suitable way. Vacuum with $\omega = 0$ in Brans-Dicke theory is described by

$$G_{ab} \equiv R_{ab} - \frac{1}{2}g_{ab}R = \tfrac{1}{\phi}\nabla_a\nabla_b\phi \equiv S_{ab}, \tag{1}$$

---

[1]The author thanks Ira Wasserman for emphasizing this point.

where $G_{ab}$ is the Einstein tensor, $R_{ab}$ the Ricci tensor, $R=0$ the scalar curvature, $g_{ab}$ the metric and $\phi$ the real Brans-Dicke scalar field. The stress-energy tensor $S_{ab}$ is trace-free, whence $\Box\phi = 0$. It thus has the property of a unique coupling to the Einstein tensor in $G_{ab} = \kappa S_{ab}$, namely $\kappa = 0, 1$ [10]. In a sense, $S_{ab}$ is a four-dimensional, scale free generalization of surface tension in three-dimensional continuum mechanics.

Soliton solutions can be found from their transition equation [11]. In seeking soliton solutions with approximate exterior Schwarzschild space-times, we shall work with the Schwarzschild line-element

$$\mathrm{d}s^2 = e^\lambda \mathrm{d}r^2 + r^2(\mathrm{d}\theta^2 + \sin^2\theta \mathrm{d}\phi^2) - e^\nu \mathrm{d}t^2. \tag{2}$$

Earlier analytic studies, by Brans and Dicke [7,3,8] and more recently by Campanelli & Lousto [4], have employed different line-elements (including isotropic spatial coordinates). The 'black hole' solutions of Campanelli & Lousto [4] are non-standard, in that their solutions have infinite horizon surface area, and their exterior space-times are 'funnels' characterized by a local minimum of area of surfaces of constant $r$.

## II. TRANSITION EQUATION

We have $S_{ab} = \frac{1}{\phi}\nabla_a\partial_b\phi = \frac{1}{\phi}(\delta_a^r\delta_b^r\frac{\phi''}{\phi'} - \Gamma_{ab}^r)\phi'$, where $\Gamma_{ab}^c$ are the connections and $f' \equiv \frac{\mathrm{d}f}{\mathrm{d}r}$. Because of $\Box\phi = 0$, we have $\frac{\phi''}{\phi'} = e^\lambda g^{ab}\Gamma_{ab}^r$ and $\phi' = -\frac{A}{r^2}e^{\frac{\lambda-\nu}{2}}$, where $A$ is an integration constant, which may be taken to be positive. Because $\Gamma_{ab}^r = \mathrm{diag}(\frac{\lambda'}{2}, -re^{-\lambda}, -r\sin^2\theta e^{-\lambda}, \frac{\nu'}{2}e^{\nu-\lambda})$, we have $\frac{\phi''}{\phi'} = e^\lambda g^{ab}\Gamma_{ab}^r = (\frac{\lambda'}{2} - \frac{\nu'}{2} - \frac{2}{r})$. In manifestly trace-free form

$$S_a^b = \mathrm{diag}(\frac{1}{2}r\nu' + 2, -1, -1, -\frac{1}{2}r\nu')\frac{e^\gamma}{r^2}, \tag{3}$$

where $e^\gamma = \frac{A}{r\phi}e^{-\frac{\lambda+\nu}{2}}$.

Einstein's equations (1) become

$$\begin{cases} R_r^{\ r} - \frac{1}{2}R = e^{-\lambda}(\frac{\nu'}{r} + \frac{1}{r^2}) - \frac{1}{r^2} = (\frac{1}{2}r\nu' + 2)\frac{e^{\gamma}}{r^2}, \\ R_\theta^{\ \theta} - \frac{1}{2}R = e^{-\lambda}(\frac{\nu''}{2} + \frac{\nu'^2}{4} - \frac{\nu'\lambda'}{4} + \frac{\nu'-\lambda'}{2r}) = -\frac{e^{\gamma}}{r^2}, \\ R_t^{\ t} - \frac{1}{2}R = -e^{-\lambda}(\frac{\lambda'}{r} - \frac{1}{r^2}) - \frac{1}{r^2} = -\frac{1}{2}r\nu'\frac{e^{\gamma}}{r^2}. \end{cases} \quad (4)$$

To obtain a differential equation for $\gamma$, write $\phi = e^{-\frac{\lambda+\nu}{2}}\psi$. Upon differentiation, we have $-\frac{A}{r^2}e^{\frac{\lambda-\nu}{2}} = \phi' = -\frac{1}{2}(\lambda'+\nu')\phi + e^{-\frac{\lambda+\nu}{2}}\psi'$, while substraction of the first and third of Einstein's equations above gives $\nu' + \lambda' = (r\nu' + 2)\frac{e^{\lambda+\gamma}}{r} = (r\nu' + 2)\frac{A}{r^2\phi}e^{\frac{\lambda-\nu}{2}}$. Combining the last two identities gives $\psi' = \frac{1}{2}r\nu'\frac{A}{r^2}e^{\lambda}$. Now, $re^{\gamma} = \frac{A}{\psi}$, whence $e^{\gamma}(1 + r\gamma') = (\frac{A}{\psi})' = -e^{\lambda}e^{2\gamma}\frac{1}{2}r\nu'$. Thus, we shall work with

$$\begin{cases} 1 + r\gamma' = -e^{\lambda+\gamma}\frac{1}{2}r\nu', \\ 1 + r\nu' = e^{\lambda}(1 + (\frac{1}{2}r\nu' + 2)e^{\gamma}), \\ 1 - r\lambda' = e^{\lambda}(1 - \frac{1}{2}r\nu'e^{\gamma}). \end{cases} \quad (5)$$

Now, introduce $z = \log\frac{r}{r_0}$, where $r_0$ is a fiducial mass. Then the combinations $rf'$ reduce to $\frac{df}{dz}$, which we shall also denote by $f'$. System (5) becomes the autonomous system of ordinary differential equations

$$\begin{cases} \gamma' + \frac{1}{2}e^{\lambda+\gamma}\nu' = -1, \\ (1 - \frac{1}{2}e^{\lambda+\gamma})\nu' = -1 + e^{\lambda} + 2e^{\lambda+\gamma}, \\ \lambda' - \frac{1}{2}e^{\lambda+\gamma}\nu' = 1 - e^{\lambda}, \end{cases} \quad (6)$$

in which $\nabla^a S_{ab} = 0$ has been used to replace the second of Einstein's equations. Note that solutions to (6) are approximately Schwarzschild, wherever $\frac{1}{2}e^{\lambda+\gamma} \ll 1$.

Next, consider $Z = \frac{1}{2}e^{\lambda+\gamma}$. System (6) gives $\lambda' + \frac{1-2Z}{1-Z}(e^{\lambda}-1) - \frac{4Z^2}{1-Z} = 0$ with $Z' + e^{\lambda}Z = 0$. Let $f(z) = e^{-\lambda(z)}$, so that

$$(1 - Z)ff' + (1 - 2Z)(f - 1)f + 4Z^2 f^2 = 0 \quad (7)$$

with $Z' + \frac{Z}{f} = 0$. Because $Z$ is a strictly monotone function (in the presence of nontrivial $S_{ab}$ and wherever the solution is defined), we are at liberty to change the independent variable: $f(z) \equiv g(Z(z))$. Note that $\frac{d}{dz}f(z) = -\frac{Z}{g}\frac{d}{dZ}g(Z)$. We shall denote $\frac{d}{dZ}g(Z)$ also by $g'$, and hence $Z(1 - Z)g' + (1 - 2Z)(1 - g)g - 4Z^2g^2$, that is,

$$\{(Z - Z^2)g\}' - g^2(4Z^2 - 2Z + 1) = 0. \tag{8}$$

Writing $h = (Z - Z^2)g$, we get $-(\frac{1}{h})' = \frac{4Z^2 - 2Z + 1}{(Z - Z^2)^2} = \frac{1}{Z^2} + \frac{3}{(1-Z)^2}$. Evaluation gives

$$e^\lambda = 1 - 4Z + \tfrac{1}{\epsilon}Z(1-Z) \equiv \tfrac{1}{\epsilon}(Z_1 + Z)(Z_2 - Z), \tag{9}$$

where $\frac{1}{\epsilon}$ is an integration constant. Consequently, (6) gives the reduced equation $\nu'(z) = \frac{1}{\epsilon}Z(z)$. The zeros $Z = -Z_1 \sim -\epsilon$ and $Z = Z_2 \sim 1 - 3\epsilon$ as $\epsilon$ becomes small. By the aforementioned relation $Z' + e^\lambda Z$, $Z(z)$ satisfies

$$Z' = \frac{1}{\epsilon}Z(Z + Z_1)(Z - Z_2), \tag{10}$$

which we shall refer to as the the *transition* equation (*cf.* the analogous, but second order equations in [11]). The solutions to the transition equation are sketched in Figure 1. With three zeros in its 'potential' $\frac{1}{\epsilon}Z(Z + Z_1)(Z - Z_2)$, the transition equation possesses two transitions for a given $\epsilon$. With the choices $\epsilon > 0$ and $\epsilon < 0$, therefore, there are altogether four different possible transitions, that is, four types of solutions. The solutions to the transition equation have a transition from either $Z = -Z_1$ or $Z = Z_2$ to the exterior Schwarzschild limit $Z = 0$ if $\epsilon > 0$, or from the interior Schwarzschild limit $Z = 0$ to either $Z = -Z_1$ or $Z = Z_2$ if $\epsilon < 0$. These solutions can be regarded to be of exterior and interior Schwarzschild type, respectively. The first, exterior Schwarzschild type solutions are the soliton solutions of interest to us.

### III. NONTOPOLOGICAL SOLITONS

To integrate the transition equation, write $Z'\{\frac{1}{Z} - \frac{1}{Z_2 + Z_1}(\frac{Z_1}{Z - Z_2} + \frac{Z_2}{Z + Z_1})\} = -\frac{1}{\epsilon}Z_1Z_2$, so that $Z(z)$ follows from the transcedental equation

$$\frac{|Z|^{Z_1 + Z_2}}{|Z_2 - Z|^{Z_1}|Z + Z_1|^{Z_2}} = Ke^{-z\frac{Z_1 Z_2}{\epsilon}(Z_1 + Z_2)} \equiv Ke^{-z\eta}, \tag{11}$$

where $K$ is an integration constant and $\eta(\epsilon) = \frac{Z_1 Z_2}{\epsilon}(Z_1 + Z_2) \sim 1$. By the aforementioned relations $\nu' = \frac{1}{\epsilon}Z$ and $Z' + e^\lambda Z = 0$, $\nu$ can be integrated to give

$$e^\nu = \Big(\frac{1 - \frac{Z}{Z_2}}{1 + \frac{Z}{Z_1}}\Big)^{\frac{1}{Z_1 + Z_2}}, \tag{12}$$

which, together with (9) and the transcedental equation (11), completely defines the line element.

The solutions can be treated analytically for the special case $\epsilon = \frac{1}{4}$, in which case the transition equation reads $Z' = 4Z(Z^2 - \frac{1}{4})$, and $Z_1 = Z_2 = \frac{1}{2}$. With the choice $K = 1$, the transcedental equation can be inverted analytically to obtain the asymptotically flat soliton solution

$$\begin{cases} e^\lambda = \frac{1}{1 + (\frac{r_0}{r})^2}, \\ e^\nu = \frac{\sqrt{1 + (\frac{r_0}{r})^2} - \frac{r_0}{r}}{\sqrt{1 + (\frac{r_0}{r})^2} + \frac{r_0}{r}}. \end{cases} \tag{13}$$

The solutions of interest are those with $Z \sim 0$ as $z \to \infty$. For small $\epsilon$, $Z_1 + Z_2 \sim 1, \eta \sim 1, Z_1 \sim \epsilon, Z_2 \sim 1$ and $(Z_2 - Z)^{Z_1} \sim 1$. If $\epsilon > 0$, but small, therefore, $\frac{Z}{Z+\epsilon} \sim Ke^{-z}$, and hence $\frac{Z}{\epsilon} \sim \frac{Ke^{-z}}{1 - Ke^{-z}}$. The approximation to the exterior Schwarzschild solution thus follows:

$$\begin{cases} e^\lambda \sim 1 + \frac{Z}{\epsilon} \sim \frac{1}{1 - Ke^{-z}} = \frac{1}{1 - \frac{2M}{r}}, \\ e^\nu \sim \frac{1}{1 + \frac{Z}{\epsilon}} \sim 1 - Ke^{-z} = 1 - \frac{2M}{r}, \end{cases} \tag{14}$$

after imposing the relationship $Kr_0 = 2M$, where $M$ the Schwarzschild mass of the hole. Because for $\epsilon < 0$ the transition equation shows soliton solutions with $Z \to Z_1 \neq 0$ as $z \to \infty$, i.e., non-Schwarzschild-like exterior space-times, $\epsilon > 0$ shall be considered henceforth. The interior space-time $(r < 2M)$ of the solitons is similarly found to be

$$\begin{cases} e^\lambda \sim \frac{1}{\epsilon}(\frac{r}{2M})^{1/\epsilon}, \\ e^\nu \sim \epsilon(\frac{r}{2M})^{1/\epsilon}, \end{cases} \tag{15}$$

Note the extremely rapid decay towards zero of both $e^\lambda$ and $e^\nu$ as $r \to 0$, while $e^{\lambda - \nu} \sim \frac{1}{\epsilon^2}(\frac{r}{2M})^{-2}$. Consequently, $\phi'(r) \sim -\frac{1}{2M\epsilon}(\frac{r}{2M})^{-3}$ as $r \to 0$ (taking $A = 2M$), so that $\phi \sim \frac{1}{2\epsilon}(\frac{r}{2M})^{-2}$. Note, therefore, that the densities $\sqrt{-g}\phi$ and

$$S_a^b \sqrt{-g} \sim 2\text{diag}(\frac{1}{2\epsilon} + 2, -1, -1, -\frac{1}{2\epsilon})\frac{r}{2M} \tag{16}$$

are finite and regular near the origin. We shall interpret the solutions as regular, nontopological soliton solutions for $Z$. The coupling to the metric is through $Z \to \phi \to S_{ab}$.

The transition represents an effective horizon in that it represents a thin layer of arbitrarily high redshift. Thus, in the classical sense, the horizon is removed by $S_{ab}$, and reduced to an effective horizon for sufficiently small $\epsilon$. In particular, with $\epsilon \sim \epsilon_P = l/2M$, where $l$ is the Planck length, the distinction between the effective horizon in the presence of $S_{ab}$ and a true horizon in the absence of it can only be discussed in the context of quantum gravity for black holes. The new solitons with small $\epsilon$ can therefore be regarded as approximations to Schwarzschild black holes, in that the difference between the exterior space-time of the solitons and that of Schwarzschild black holes is within the realm of quantum gravity for $\epsilon \sim \epsilon_P$. We expect that the present arguments carry through in the case of Kerr black holes, but have not pursued this point.

If $S_{ab}$ is indeed physical, the extreme soliton solutions put Hawking's theorem [9] on the non-existence of black holes in BD theory with non-constant $\phi$ in somewhat different perspective. Hawking's theorem is purely classical and does not take into account (nor prohibit) the possibility encountered here of non-black hole, soliton solutions which are classically indistinguishable from black holes, their difference being within the realm of quantum gravity. Their *interior* space-times, however, are dramatically distinct from that in standard black holes. In this context, note that a Schwarzschild black hole remains a solution of (1), by considering $\phi =$const, so that, in fact, (1) defines two solutions with like exterior space-times. The nontopological solitons, therefore, may be considered bifurcations from Schwarzschild black holes in space-time with $S_{ab}$, where the bifurcation parameter is $\epsilon$.

## IV. CONCLUSIONS

The approximation of black holes by extreme soliton solutions offers the advantages of a global time-like coordinate and a regularization of horizon surfaces due to an everywhere finite (but arbitrary large) redshift (and hence allowing for a great degree of freedom in space-time slicing). In light of this, the solitons are proposed as an *Ansatz* for treating black holes in numerical relativity. Perhaps the *a priori* most immediate aspects to be addressed are (i) the nature and handling of the singularity at the origin and (ii) the persistence of the soliton solutions during genuine three-dimensional time-evolutions. It would be of interest to consider the effect of a finite coupling constant, $\omega << 1$, on the behavior of $\phi$ near the origin. The strong interaction of two solitons in a binary system, furthermore, challenges their existence, in that radiation of $\phi$ to infinity may threaten to disperse the solitons. It should be appreciated that small $\epsilon$ serves to reduce radiation of $\phi$ from the region near the effective horizon, in view of the associated high redshift. It would take three-dimensional experiments to verify these radiation and stability aspects.

**Acknowledgment.** The author greatfully acknowledges stimulating discussions with Amir Levinson, Ira Wasserman, and Saul Teukolsky. This work has been supported in part by NSF grant 94-08378 and by the Grand Challenge grant NSF PHY 93-18152/ASC 93-18152 (ARPA supplemented). The Cornell Theory Center is supported by NSF, NY State, ARPA, NIH, IBM and others.

# REFERENCES

[1] Blanchet L., Damour T., Iyer B., Will C.M. & Wiseman A.G. Gravitational-radiation damping of compact binary systems to second post-newtonian order. *Phys. Rev. Lett.*, 74(18):3515–3518, 1995.

[2] Blanchet L., Damour T. & Iyer B. Gravitational waves from inspiralling compact binaries: energy loss and waveform to second-post-newtonian order. *Phys. Rev. D.*, pages 5360–5386, 1995.

[3] Brans C. *Physical Review*, 125(6):2194–2201, 1962.

[4] Campanelli M. & Lousto C.O. *Int. J. Mod. Phys. D.*, 2(4):451–462, 1993.

[5] Bradaschia C. Caloni E. Cobal M. Del Fasbro R. Di Virgilio A. Giazotta A. Holloway E. Kautzky H. Michelozzi B. Montelatici V. Pascuello D. and Proceedings of the Banff Summer Institute, Banff, Alberta, edited by R. Mann & P. Wesson (World Scientific, Singapore, 1991). Velloso W. in *Gravitation 1990*.

[6] Abramovici A., Althouse W.E., Drever R.W.P., Gursel Y., Kawamura S., Raab F.J., Shoemakes D., Siewers L., Spero R.E., Thorne K.S., Vogt R.E., Weis R., Whitcomb S.E. & Zucker M.E. *Science*, 256:325, 1992.

[7] Brans C. & Dicke R.H. *Physical Review*, 124(3):925–935, 1961.

[8] Dicke R.H. *Physical review*, 125(6):2163–2167, 1962.

[9] Hawking S.W. *Comm. Math. Phys.*, 25:167–171, 1972.

[10] van Putten H.P.M. *Lett. Math. Phys.*, (submitted).

[11] Lee T.D. & Pang Y. *Physics Reports*, 221(5-6):251–350, 1992.

**FIGURE 1.** The solutions to the transition equation $Z' = \frac{1}{\epsilon}Z(Z+Z_1)(Z-Z_2)$ for $\epsilon > 0$. The curves shown are those for $\epsilon = \frac{1}{4}$; for general $\epsilon > 0$ the asymptotes $Z_1 \neq Z_2$. Two nontopological soliton solutions are shown, corresponding to the transition Ia ($Z \sim Z_2$ to $Z \sim 0$) and Ib ($Z \sim -Z_1$ to $Z \sim 0$. The two diverging solutions IIa and IIb describe finite volume solutions, existing for $r < r_0$ only, while the two constant solutions IIIa ($Z = Z_2$) and IIIb ($Z = -Z_1$) solutions do not approach the exterior Schwarzschild solution $Z = 0$. The soliton solution Ia is the nontopological soliton of interest.

**FIGURE 2.** The nontopological soliton for the case $\epsilon = 1/100$. A transition layer forms at $r = r_0$ with high redshift as $\epsilon$ becomes small. The transition layer serves to replace the horizon of the correpsonding Schwarzschild black hole. The radius is normalized with respect to $2M$. The exterior space-time of the soliton approximates the exterior Schwarzschild space-time.

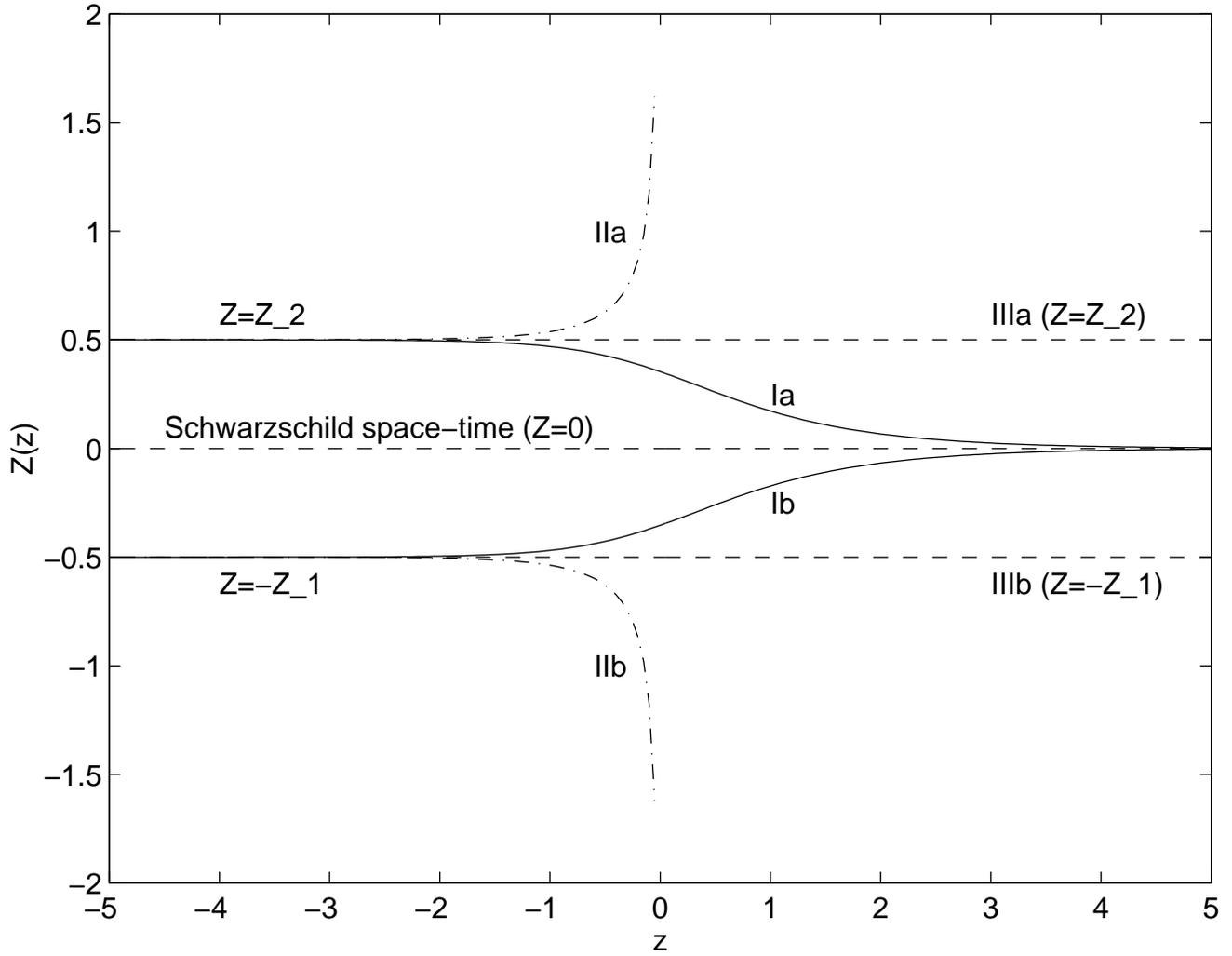

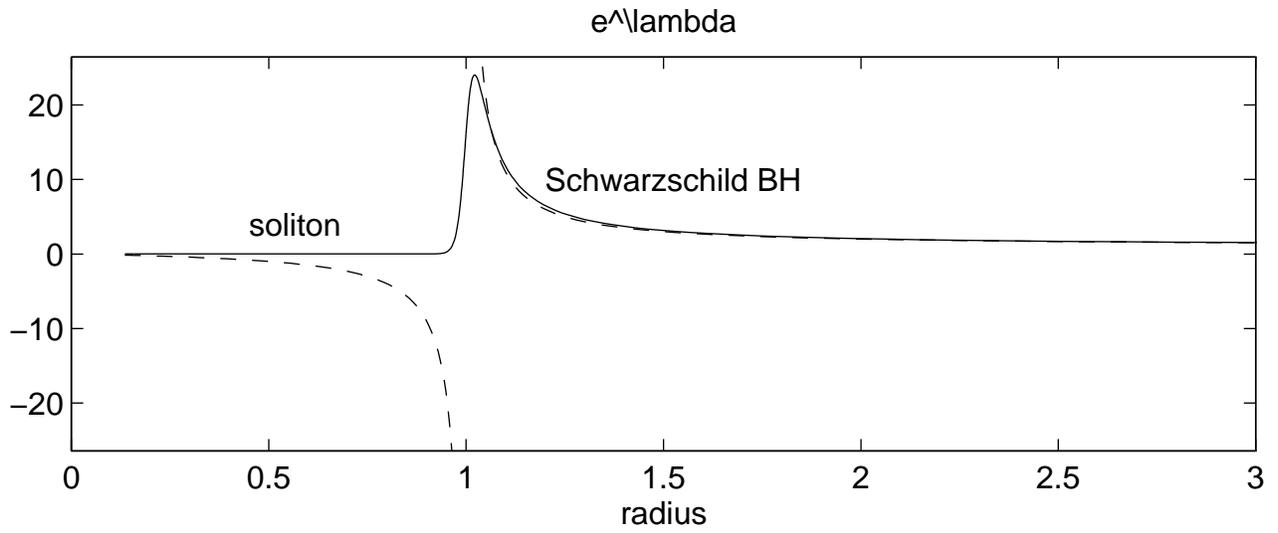

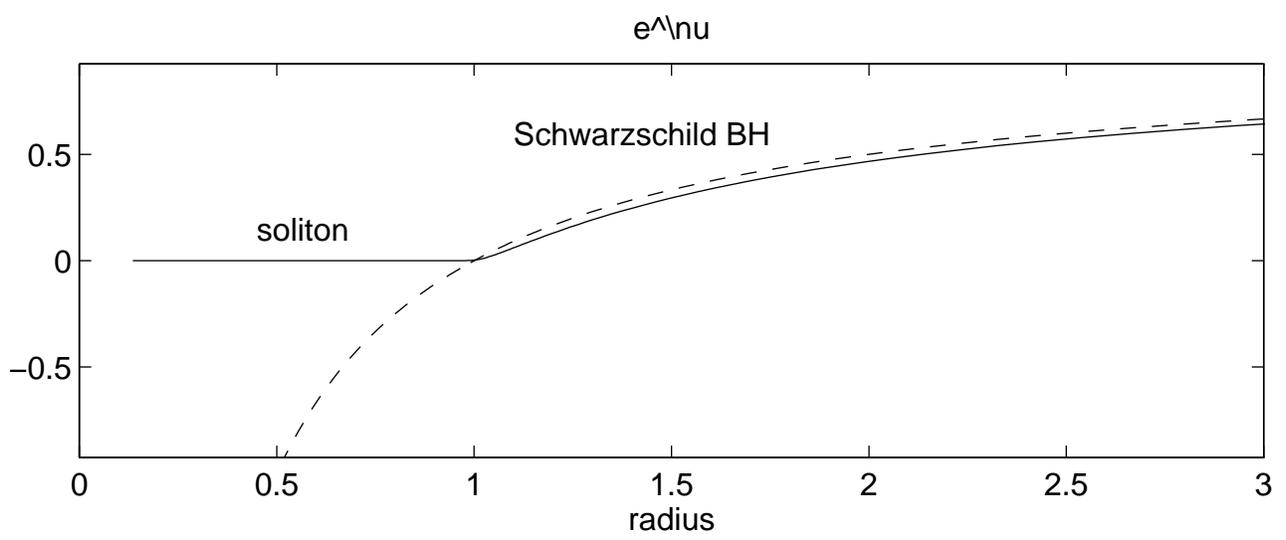